# Phase-amplitude functional theory – new ab initio calculation method for large size systems

Pawel Strak[1], Konrad Sakowski[1,2,3], Pawel Kempisty[1,3] and Stanislaw Krukowski[1]

[1]*Institute of High Pressure Physics, Polish Academy of Sciences, Sokolowska 29/37, 01-142 Warsaw, Poland*

[2]*Institute of Applied Mathematics and Mechanics, University of Warsaw, Banacha 2, 02-097 Warsaw, Poland*

[3]*Research Institute for Applied Mechanics, Kyushu University, Kasuga, Fukuoka 816-8580, Japan*

New method for *ab initio* calculations of the properties of large size system based on phase-amplitude functional is presented. It is shown that Schrodinger equation for many electrons complex system including large size molecules, or clusters and also periodic systems could be translated into functional of two variables, attributed to many electron wavefunctions: phase and the amplitude (i.e. square root of total electron density). The equations for the phase and the amplitude are derived. The kinetic and Coulomb interaction energy are expressed in function of these variables. The equations for one-electron wavefunctions, necessary for the energy spectrum are derived using these two variables.

# I. INTRODUCTION

Derivation of density functional theory basic equations by Kohn and Sham did not lead to immediate substantial progress in *ab intio* simulations of the properties of complex system by quantum mechanical methods [1]. At that time Hartree-Fock (HF) method dominated ab initio simulation of small scale systems due to its relatively high precision and numerical implementation in simulation packages [2,3]. The method was limited to very small systems as its requirements excluded application to larger size systems.

Hartree-Fock method, served as reference point for development of several trends in *ab initio* simulations. On one hand, the precision of HF method was not satisfactory for determination of such properties as interaction energy vs distance dependence for small molecules. The precision was to be higher than $10^{-5}$ of the total energy that requires considerable improvement of the simulations. That led to creation of post-Hartree-Fock methods, such as configuration interaction (CI) [4,5] or coupled cluster singles, doubles and perturbation triples (CCSD(T)) as most successful examples of this trend [6-8].

The second trend was towards extension of the size of the system, even at some deterioration of the precision of the results. This originated from tight-binding and effective mass approximation to understanding of the properties of solids. That brought qualitative understanding of the basic properties of solids, such as metals, semiconductors, etc. This was achieved by partially phenomenological approach to adjust the results to experimental data. The results were far from satisfactory. Despite that these approaches are used also recently, e.g. in the investigations of graphene properties effective mass approximation played dominant role.

A most substantial progress was made by the development of density functional theory. The original set of equation was subsequently supplemented by better parametrization of the exchange-correlation terms [9]. Considerable improvements was attained by introduction of pseudopotential that allowed to simulate much larger systems by relatively small deterioration of the precision of the calculations [10,11]. Thus, the system sizes are increased allowing to obtain results of great physical importance. Nevertheless the limitation remains severe. Recent improvement are not significant, most progress is due to better computer equipment, parallelization of the algorithm and correction schemes.

The absence of substantial progress is related to the fact that all modifications are within the basic scheme of DFT formalism [12-17]. Thus density functional is employed and the pseudopotentials are used. Further considerable progress can be made by drastic change of the basic formalism. The present paper is such an attempt, the density functional is replaced by double: amplitude and phase functionals. Which allow to express the kinetic energy also with this framework.

## II. THE FUNCTIONAL FORMALISM

### 1. Stability equations

It is proposed that multi-electron wavefunction is expressed as

$$\Psi_{A\Phi} = A(r)e^{i\Phi(r)} \tag{1}$$

where the amplitude $A(r)$ and the phase $\Phi(r)$ are positively defined real functions in the three-dimensional space. The amplitude $A(r)$ is related to the electron density in standard way: $\rho(\vec{r}) = |A(\vec{r})|^2$.

The expressions for the energy contribution were derived in appendices A, B and C. The total Hamiltonian is expressed as the sum:

$$H = T[A,\varphi] + U_{XC}[A,\varphi] + U_{Coul}[A] + U_{e-i}[A] + V_{ext}[A] \tag{2}$$

i.e. the kinetic and exchange energy depends on the amplitude and the phase while the Coulomb electron-electron and also electron-ion interaction and external energies on the amplitude only. This division stems from the fact that kinetic and exchange energies depend on the electron momentum while the interaction with ions and external fields depends on the electron location only. As the amplitude and the density of electrons are linked by standard relationship, given by Eq. A12, the variation should not change the number of the electrons:

$$\int \rho(\vec{r}) d^3r = \int A^2(\vec{r}) d^3r = N \tag{3}$$

The stationary equations of motion require that the system does not evolve in time, i.e. it is in energy minimum with respect to the amplitude and the phase:

$$\frac{\delta H[A,\varphi]}{\delta A}(\vec{r}) = EA(\vec{r}) \tag{4a}$$

$$\frac{\delta H[A,\varphi]}{\delta \varphi}(\vec{r}) = 0 \tag{4b}$$

The E parameter is Lagrange factor arising from the conservation of the total number of electrons $N$ could be interpreted as the system total energy. The functional derivatives of the Hamiltonian components in Eq. 1 could be obtained, giving

- For the kinetic energy

$$\frac{\delta T[A,\varphi]}{\delta A}(\vec{r}) = \frac{\hbar^2 S}{2m}\{2\nabla^2 A(\vec{r}) - 2A(\vec{r})(\nabla\phi(\vec{r}))^2\} \tag{5a}$$

$$\frac{\delta T[A,\varphi]}{\delta \varphi}(\vec{r}) = \frac{\hbar^2 S}{2m}\{2\nabla(A(\vec{r})\nabla\varphi(\vec{r}))\} \tag{5b}$$

- For the Coulomb energy

$$\frac{\delta U_{Coul}[A]}{\delta A}(\vec{r}) = 4A(\vec{r}) \int d^3r' \frac{e^2 A^2(\vec{r}')}{4\pi\varepsilon_o|\vec{r}-\vec{r}'|} \tag{6a}$$

$$\frac{\delta U_{Coul}[A]}{\delta \varphi}(\vec{r}) = 0 \tag{6b}$$

- For the exchange energy

$$\frac{\delta U_{XC}[A,\varphi]}{\delta A}(\vec{r}) = 4A(\vec{r}) \int d^3r' \frac{e^2 A^2(\vec{r}\prime)\cos[2(\varphi(\vec{r})-\varphi(\vec{r}\prime))]}{4\pi\varepsilon_o|\vec{r}-\vec{r}\prime|} \tag{7a}$$

$$\frac{\delta U_{XC}[A]}{\delta \varphi}(\vec{r}) = -4A^2(\vec{r}) \int d^3r' \frac{e^2 A^2(\vec{r}\prime)\sin[2(\varphi(\vec{r})-\varphi(\vec{r}\prime))]}{4\pi\varepsilon_o|\vec{r}-\vec{r}\prime|} \tag{7b}$$

- For the electron-ion interaction energy

$$\frac{\delta U_{e-i}[A]}{\delta A}(\vec{r}) = \sum_\alpha \left( \frac{2Z_\alpha e^2 A(\vec{r})}{4\pi\varepsilon_o|\vec{R}_\alpha-\vec{r}|} \right) \tag{8a}$$

$$\frac{\delta U_{e-i}[A]}{\delta \varphi}(\vec{r}) = 0 \tag{8b}$$

- For the external field energy

$$\frac{\delta V_{ext}[A]}{\delta A}(\vec{r}) = 2A(\vec{r})V_{ext}(\vec{r}) \tag{9a}$$

$$\frac{\delta V_{ext}[A]}{\delta \varphi}(\vec{r}) = 0 \tag{9b}$$

These variations are used via Eqs 3a and 3b to obtain variational equations for stability of the system which gives

$$\frac{\hbar^2 S}{2m}\{2\nabla^2 A(\vec{r}) - 2A(\vec{r})(\nabla\phi(\vec{r}))^2\} + 4A(\vec{r}) \int d^3r' \frac{e^2 A^2(\vec{r}\prime)\{1-\cos[2(\varphi(\vec{r})-\varphi(\vec{r}\prime))]\}}{4\pi\varepsilon_o|\vec{r}-\vec{r}\prime|} + \sum_\alpha \left(\frac{2Z_\alpha e^2 A(\vec{r})}{4\pi\varepsilon_o|\vec{R}_\alpha-\vec{r}|}\right) =$$
$$EA(\vec{r}) \tag{10a}$$

$$\frac{\hbar^2 S}{2m}\{2\nabla(A(\vec{r})\nabla\varphi(\vec{r}))\} - 4A^2(\vec{r}) \int d^3r' \frac{e^2 A^2(\vec{r}\prime)\sin[2(\varphi(\vec{r})-\varphi(\vec{r}\prime))]}{4\pi\varepsilon_o|\vec{r}-\vec{r}\prime|} = 0 \tag{10b}$$

These equations have to be solved using appropriate boundary conditions for two real, positive defined functions: $A(\vec{r}) > 0$ and $\varphi(\vec{r}) > 0$.

## 2. Boundary conditions

The boundary conditions have to be specified for two different types of the systems:

(i) the periodic system, used for simulation of the infinite system, predominantly the crystals

(ii) the finite systems, used for simulations of atoms, molecules and clusters

The boundary conditions for the amplitude are essentially derived from the analogous conditions used in density functional theory, as the amplitude and the density are directly related. Therefore the boundary conditions for the amplitude have to reflect periodicity of the physical property X of the system with respect to translation by vector $\vec{L}$:

$$X(\vec{R}) = X(\vec{R} + \vec{L}) \tag{11}$$

As the amplitude (electron density) determines the physical properties of the system, the condition is analogous:

$$A(\vec{R}) = A(\vec{R} + \vec{L}) \tag{12}$$

The condition for the phase is different, as the phase is not so directly related which is reflected by the Bloch function, being the solution for wavefunction in the periodic systems, e.g. crystals. There the analogy between the Bloch function and the free electron theory should be expressed by the boundary conditions for the latter. Assuming that the phase is positively defined, the condition is (for 1D):

$$\Delta\varphi(L) = 2\pi(2\sum_{n=1}^{M} n) = 2\pi M(M+1) \tag{13}$$

where $\Delta\varphi(L)$ is the phase difference across the sample of the length L, M is the maximal occupied number and factor of 2 stems from positive sign of the phase. For 3 D the condition for number of valence electron N is therefore:

$$3(2\sum_{n=1}^{M} n) = 3M(M+1) = \frac{N}{2} \tag{13}$$

from which the maximal number M is readily obtained:

$$M = \frac{1}{2}\left[\sqrt{\frac{2N+3}{3}} - 1\right] \tag{14}$$

In 3 d these difference should be enforced in all three directions. In case M is not integer number, the conditions are using fraction as the fractional occupation may be allowed. Note that the phase amplitude has to be enforced in order to avoid the trivial solution $\varphi(\vec{r}) = 0$ for which the kinetic energy disappears. Such solution will give zero kinetic energy contribution, and accordingly, it would violate fermionic character of the electron set.

In case of the finite size systems, again the condition for amplitude follows that of density in DFT theory, and more precisely, at far distance the amplitude should vanish in exponential way:

$$A(\vec{R}) = A(L) \propto \frac{exp(-\alpha L) \to 0}{L \to \infty} \tag{15}$$

in accordance to the general linear character of the theory. For the phase the boundary condition may be derived from simple solution of Schrodinger equation for single atom as the phase should follow the long distance behavior of single potential center. Thus it should behave as free electron with respect of the angle, i.e. should be linear. Naturally, the phase should be periodic over full angle, that enforced the same limitation as for periodic system. Thus Eq. 13 apply.

### 3. Eigenfunctions

The eigenfunctions of the considered system may be obtained by several procedures. The most obvious way is to use linearized version of Kohn-Sham equation, using amplitude and density relationship in Eq. A12 [9]. Thus, the linear solution of Kohn-Sham equation has to be solved, which effectively reduces self-consistent field (SCF) iteration procedure to single one. This brings considerable reduction of the computational resourced needed to solve the problem. In addition, the use of the consistent algebraic matrix greatly reduces the probability of divergence of the procedure.

A second possible approach is to use Kohn-Sham equations for the variations with respect to the change of the wavefunctions [9]:

$$\frac{\delta H[\rho]}{\delta \psi_i}(\vec{r}) = \varepsilon_i \psi_i(\vec{r}) \quad (15)$$

where the eigenfunctions $\psi_i$ are normalized:

$$\int |\psi_i(\vec{r})|^2 \, d^3r = 1 \quad (16)$$

and using relations between the electron density (amplitude), the phase and the single electron wavefunctions that gives

$$\frac{\delta H[A,\varphi]}{\delta A} \frac{\delta A}{\delta \psi_i}(\vec{r}) + \frac{\delta H[A,\varphi]}{\delta \varphi} \frac{\delta \varphi}{\delta \psi_i} = \varepsilon_i \psi_i(\vec{r}) \quad (17)$$

that could be resolved using Eqs. 10.

Finally, it is worth to account that all these procedures have to express the single electron wavefunctions in series of linear combinations of basis functions belonging to the predetermined function sets, such as plane waves (e.g. VASP [15,16]), atomic or molecular orbitals (e.g. SIESTA[12,13]), or wavelets (e.g. BiGDFT [18]). The series tend to be bulky, thus these functions are prone to generate extremely large projection matrices. Therefore only relatively small size of the system could be represented. The adopted solution is to reduce the represented function by the use of pseudopotentials and to reduce the basis sets. The procedure may be applied to these equations again in the same way. It is worth to say that generally, the procedure still produces oversized set which should be reduced further. These procedures has to be developed in the future [19]

### III. THE EXAMPLES OF APPLICATIONS

#### 1. Free non-interacting electron system – stationary state

The free electron system is the example used for the formulation of the code. Nevertheless it is useful to find whether such solution fulfills the assumptions of the theory. Assume that we have periodic system in a box of the size L, having G electrons. The solution is:

$$\Psi_{A\Phi} = A(r)e^{i\Phi(r)} = A \, exp[i(\alpha_1 x + \alpha_2 y + \alpha_3 z)] \quad (19)$$

which has to be verified. As the electrons do not interact, Eq. 10a is reduced to kinetic energy terms in which only the second term is not zero, i.e.

$$\frac{\hbar^2 S}{m}(\alpha_1 + \alpha_2 + \alpha_3)^2 = E \quad (20)$$

which is fulfilled due to definition of S factor. For the second equation, the differentiation gives

$$\nabla \varphi(\vec{r}) = [\alpha_1, \alpha_2, \alpha_3] = const \qquad (21)$$

i.e. constant vector which upon second differentiation gives zero thus fulfilling Eq. 10 b. Therefore the equations are in agreement with the proposed function form.

## IV. SUMMARY

Stability equations using phase-amplitude functional space were derived for set of fermion systems, with the emphasis for application of multi-electron system typical for large scale molecules and clusters and also to infinite system in slab geometry. This provides set of two equations sufficient to obtain basic stability of the large size systems. These equations were supplemented with the definition of the boundary conditions for the phase and the amplitude that is equivalent to density. The basic difference with traditional density functional equation lies in the formulation of kinetic energy term, which is scaled to obtain proper kinetic energy value from the phase component.

After solution of the basic equations, the extended set of equations regarding one-electron wavefunctions has to be solved, either using standard Kohn-Sham equations in linearized form, or newly derived equations. This could be done to entire system, or if the size is prohibitive, to the part of it. This part has to be developed further to attain a possibility of ab initio calculations for the system sizes comprising large number of atoms, typical for nanotechnology applications. In the more distant perspective, this opens route to more efficient progress in the intensively developed field of nanotechnology.

**Acknowledgements**

The research was partially supported by Polish National Science Centre grants number DEC-2015/19/B/ST5/02136 and 2017/27/B/ST3/01899. This research was carried out with the support of the Interdisciplinary Centre for Mathematical and Computational Modelling at the University of Warsaw (ICM UW) under grant no G15-9.

## Appendix A. Kinetic energy functional

Kinetic energy formulation in standard is the main problem in the DFT formulation. This is based on the mere fact that, the basis of the DFT theory is free electron solution. In this case, the solution of the Schrodinger equation is trivial, that is plane wave:

$$\psi(\vec{r}_1, t) = a\, exp[i(\vec{k}_1\vec{r}_1 - \omega t)] \tag{A1}$$

where $a$ denotes amplitude such that the wavefunction norm is set to unity, $\vec{k}$ and $\omega$ are wavevector and frequency, associated with the momentum $(\vec{p} = \hbar\vec{k})$, and the energy $(E = \hbar\omega)$, respectively. In the simple case presented here, the potential energy is absent. The kinetic energy is readily obtained by application of kinetic energy operator to this state

$$T = \langle\psi|T|\psi\rangle = \left\langle\psi\left|-\frac{\hbar^2\nabla^2}{2m}\right|\psi\right\rangle = \frac{\hbar^2 k^2}{2m} \tag{A2}$$

Most importantly, in this formulation, the one-electron wavefunction can be expressed using two basic parameters: a – amplitude and $\varphi$ - phase:

$$\psi(\vec{r}, t) = a\, exp[i(\varphi(\vec{r}) - \omega t)] \tag{A3}$$

where the phase $\varphi(\vec{r}) = \vec{k}\vec{r}$, is associated with the position of the space. That is compatible with the kinetic energy operator, acting in real space:

$$T = \langle\psi|T|\psi\rangle = \left\langle\psi\left|-\frac{\hbar^2\nabla^2}{2m}\right|\psi\right\rangle \tag{A4}$$

For two electron state the natural extension of Eq. A1 is:

$$\psi(\vec{r}, t) = a\, exp[i(\vec{k}_1\vec{r}_1 + \vec{k}_2\vec{r}_2 - \omega t)] \tag{A5}$$

that causes problems as the natural extension $\varphi(\vec{r}) = \vec{k}_1\vec{r} + \vec{k}_2\vec{r}$ leads to

$$\psi(\vec{r}, t) = a\, exp[i(\vec{k}_1\vec{r} + \vec{k}_2\vec{r} - \omega t)] \tag{A6}$$

for two electron moving in opposite directions $\vec{k}_1 = -\vec{k}_2$ would lead to cancellation of the phase $\varphi(\vec{r}) = const$ and no contribution to kinetic energy while in reality the kinetic energy is added. Thus such identification of the phase is not useful. The natural remedy is to use the absolute value, i.e.

$$\varphi(\vec{r}) = |\vec{k}_1\vec{r} + \vec{k}_2\vec{r}| \tag{A7}$$

and use positively defined phase. Naturally, the overall change of the sign of the phase does not influence the results, it reflects invariance with respect of inflection symmetry. For multi-electron (N electron) gas the solution is direct extension:

$$\psi(\vec{r}_1, \vec{r}_2, \ldots \vec{r}_N, t) = a\, exp[i(\vec{k}_1\vec{r}_1 + \vec{k}_2\vec{r}_2 + \cdots + \vec{k}_N\vec{r}_N - \omega t)] \tag{A8}$$

Asymmetry requirement changes these function to

$$\psi(\vec{r}_1, \vec{r}_2, \ldots \vec{r}_N, t) = A \sum exp[i(\vec{k}_{\sigma(1)}\vec{r}_1 + \vec{k}_{\sigma(2)}\vec{r}_2 + \cdots + \vec{k}_{\sigma(N)}\vec{r}_N - \omega t)] \tag{A9}$$

Where A is the amplitude of the multi-electron function. The sum runs over all permutations of N electrons. Natural extension of the one-electron characteristics is reformulation of multi-electron function in terms of the amplitude:

$$A = \frac{a}{\sqrt{N!}} \tag{A10}$$

and the phase

$$\Phi = |\vec{k}_{\sigma(1)}\vec{r}_1 + \vec{k}_{\sigma(2)}\vec{r}_2 + \cdots + \vec{k}_{\sigma(N)}\vec{r}_N| \tag{A11}$$

Naturally, the phase does not depend on the permutations in Eq. A3 and the density is uniform across the entire plane-wave free electron system. Reduction of the full wavefunction to the electron density, or amplitude:

$$\rho(r) \equiv |\langle\psi|\psi\rangle| = |A|^2 \tag{A12}$$

sets the relation between all electron density $\rho$ and the amplitude $A$ of multi-electron wavefunction. The dependence on the phase is totally lost. As the kinetic energy of such system is given by

$$\langle\psi|T|\psi\rangle = \sum_i \frac{\hbar^2 k_i^2}{2m} \tag{A13}$$

the kinetic energy has to be recovered from the multi-electron functional phase. That is absent in DFT formulation and it is the principal problem of the DFT theory as it is based on electron density and therefore fails to represent the kinetic energy as the density functional and has to resort to one-electron wavefunction representation.

As it was shown above the relation between the phase and the kinetic energy is not straightforward. Generally, application of kinetic energy operator $\hat{T} = -\frac{\hbar^2 \nabla^2}{2m}$ to the multi-electron wavefunction (Eq. 1) $\Psi_{A\Phi} = A(r)e^{i\Phi(r)} = A\,exp[i|\sum k_i r_i|]$ gives the following result:

$$\langle\Psi_{A\Phi}|T|\Psi_{A\Phi}\rangle = \frac{1}{2m}\left(\sum_i |\hbar k_i|\right)^2 \tag{A14}$$

which is grossly inaccurate. The kinetic energy contribution has to be rescaled to obtain the proper expression. Assume the case of 1D system of the length L, then the kinetic momentum eigenvalues are $k_i = \frac{2\pi}{L} n_i$ where $n_i = 0, \pm 1, \pm 2, ... \pm M$. Any state is occupied by the two electrons of different spin orientations as they are fermions. Thus the maximum number $M$ is related to total number of electrons $N$ in the following way: $2(2M+1) = N$. Therefore the kinetic energy obtained from Eq. A7 is:

$$\langle\psi|T|\psi\rangle = \frac{2h^2}{2mL^2}\sum_{n=-M}^{n=M} n^2 = \frac{h^2}{2mL^2}\frac{M(M+1)(2M+1)}{3} \tag{A15}$$

where we used the relation

$$\sum_{n=1}^{M} n^2 = \frac{M(M+1)(2M+1)}{6} \tag{A16}$$

Analogously it could be shown that the kinetic energy operator calculated over the phase-amplitude wavefunction gives

$$\langle\Psi_{A\Phi}|T|\Psi_{A\Phi}\rangle = \frac{2h^2}{2mL^2}\left(\sum_{n=-M}^{n=M} n\right)^2 = \frac{h^2}{2mL^2}\frac{M^2(M+1)^2}{2} \tag{A17}$$

where again the following relation was used

$$\sum_{n=1}^{M} n = \frac{M(M+1)}{2} \tag{A18}$$

Therefore the ratio of these two quantities defines the scaling factor to be used for the kinetic energy obtained from phase-amplitude wavefunction

$$S \equiv \frac{\langle\psi|T|\psi\rangle}{\langle\Psi_{A\Phi}|T|\Psi_{A\Phi}\rangle} = \frac{2(2M+1)}{3M(M+1)} \tag{A19}$$

These scaling have to be supplemented by the relation between the maximal number of the occupied state $M$ and the total number of valence electrons $N$:

$$N = 4M + 2 \tag{A20}$$

in which the two spin orientation were taken into account. Using this relation the scaling factor is:

$$S \equiv \frac{\langle\psi|T|\psi\rangle}{\langle\Psi_{A\Phi}|T|\Psi_{A\Phi}\rangle} = \frac{16N}{3(N^2-4)} \tag{A21}$$

The above scaling factor is valid for one dimension, i.e. D = 1. The *ab initio* calculations are conducted in 3D where the number of the states is different. The kinetic energy is given by

$$\langle\psi|T|\psi\rangle = \sum\frac{\hbar^2(k_x^2 + k_y^2 + k_z^2)}{2m} = \frac{h^2}{2mL^2}\sum n_x^2 + n_y^2 + n_z^2 \tag{A22}$$

where $L = L_x = L_y = L_z$. The continuous model will be used in which the maximal energy of occupied states is obtained from:

$$T = \frac{h^2}{2mL^2}\int(n_x^2 + n_y^2 + n_z^2)d^3n = 4\pi\int_0^M n^4 dn = \frac{4\pi}{5}M^5 \tag{A23}$$

where the maximal energy value $M$ is determined from the number of valence electrons $N$:

$$N = 2\int d^3n = 8\pi\int_0^M n^2 dn = \frac{8\pi}{3}M^5 \tag{A24}$$

in which the two spin orientations were taken into account. The phase-amplitude kinetic energy is then

$$\langle\Psi_{A\Phi}|T|\Psi_{A\Phi}\rangle = \frac{h^2}{2mL^2}\int(n_x + n_y + n_z)^2 d^3n = \frac{4(\pi+2)}{5}M^5 \tag{A25}$$

Thus the scaling factor in 3D is:
$$S = \frac{\pi}{\pi+2} \quad (A26)$$

The phase amplitude kinetic energy functional is therefore equal to

$$T[A,\varphi] = -\frac{\hbar^2 \pi}{(\pi+2)m} \int d^3r [A(r)^{-i\varphi} \nabla^2 (A(r)e^{i\varphi})] = -\frac{\hbar^2 \pi}{(\pi+2)m} \int d^3r \left[\sqrt{\rho(r)}^{-i\varphi} \nabla^2 (\sqrt{\rho(r)} e^{i\varphi})\right] \quad (A27)$$

which will be used in variational procedure.

## Appendix B. Electron-electron Coulomb and exchange energy functional

The electron-electron interaction contributes the two-body term to full energy functional:

$$\langle U_{e-e} \rangle = \frac{1}{2} \int d^3r_1 .. \int d^3r_N \left[\psi^*(\vec{r}_1,\vec{r}_2,...\vec{r}_N,t) \sum_{i \neq j} \left(\frac{e^2}{4\pi\varepsilon_o |\vec{r}_i - \vec{r}_j|}\right) \psi(\vec{r}_1,\vec{r}_2,...\vec{r}_N,t)\right] \quad (B1)$$

The factor in front stems from double summing pairs of electrons. The wavefunction is the antisymmetric wavefunction obtained as Slater determinant of one-electron wavefunctions. The order of integration may be changed and subsequently the order within the Slater determinant [2–9]. The dummy indices may be changed so that the only contribution stems from the integration over the variables in the interaction terms. The other wavefunctions are averaged out, leaving the two-body integration term:

$$\langle U_{e-e} \rangle = \frac{1}{2} \int d^3r_1 \int d^3r_2 \, [\psi_1^*(\vec{r}_1)\psi_2^*(\vec{r}_2) - \psi_2^*(\vec{r}_1)\psi_1^*(\vec{r}_2)] \left(\frac{e^2}{4\pi\varepsilon_o |\vec{r}_1 - \vec{r}_2|}\right) [\psi_1(\vec{r}_1)\psi_2(\vec{r}_2) - \psi_2(\vec{r}_1)\psi_1(\vec{r}_2)] \quad (B2)$$

which could be separated into the Coulomb, density dependent term

$$\langle U_{Coul} \rangle = \int d^3r_1 \int d^3r_2 \, \rho(\vec{r}_1)\rho(\vec{r}_2) \left(\frac{e^2}{4\pi\varepsilon_o |\vec{r}_1 - \vec{r}_2|}\right) \quad (B3)$$

that can be expressed as amplitude functional

$$\langle U_{Coul}[A] \rangle = \int d^3r_1 \int d^3r_2 \, A^2(\vec{r}_1)A^2(\vec{r}_2) \left(\frac{e^2}{4\pi\varepsilon_o |\vec{r}_1 - \vec{r}_2|}\right) \quad (B4)$$

and the more complex exchange term

$$\langle U_{XC} \rangle = \int d^3r_1 \int d^3r_2 \, \psi_1^*(\vec{r}_1)\psi_2^*(\vec{r}_2) \left(\frac{e^2}{4\pi\varepsilon_o |\vec{r}_1 - \vec{r}_2|}\right) \psi_2(\vec{r}_1)\psi_1(\vec{r}_2) \quad (B5)$$

which could be expressed as

$$\langle U_{XC} \rangle = \int d^3r_1 \int d^3r_2 \, \rho(\vec{r}_1)\rho(\vec{r}_2) \cos[2(\varphi(\vec{r}_1) - \varphi(\vec{r}_2))] \left(\frac{e^2}{4\pi\varepsilon_o |\vec{r}_1 - \vec{r}_2|}\right) \quad (B6)$$

or equivalently

$$\langle U_{XC}[A,\varphi] \rangle = \int d^3r_1 \int d^3r_2 \, A^2(\vec{r}_1)A^2(\vec{r}_2) \cos[2(\varphi(\vec{r}_1) - \varphi(\vec{r}_2))] \left(\frac{e^2}{4\pi\varepsilon_o |\vec{r}_1 - \vec{r}_2|}\right) \quad (B7)$$

## Appendix C. Electron-ion (e-i) interaction energy functional

The electron-ion interaction contributes the one-body term to full energy functional:

$$\langle U_{e-i} \rangle = \int d^3r_1 .. \int d^3r_N \left[\psi^*(\vec{r}_1,\vec{r}_2,...\vec{r}_N,t) \sum_\alpha \sum_i \left(\frac{Z_\alpha e^2}{4\pi\varepsilon_o |\vec{R}_\alpha - \vec{r}_i|}\right) \psi(\vec{r}_1,\vec{r}_2,...\vec{r}_N,t)\right] \quad (C1)$$

where the summation runs over all ions ($\alpha$) and electrons ($i$). $Z_\alpha$ is the atomic number of the ion ($\alpha$). Again, the order of integration and the one-electron wavefunction may be interchanged in Slater determinant, that leads to final expression for the e-i interaction

$$\langle U_{e-i} \rangle = \int d^3r_1 \left[\sum_\alpha \left(\frac{Z_\alpha e^2 \rho(\vec{r}_1)}{4\pi\varepsilon_o |\vec{R}_\alpha - \vec{r}_i|}\right)\right] \quad (C2)$$

The e-i energy does not depend on the phase, it is dependent on the density (i.e. amplitude) only.

$$\langle U_{e-i}[A] \rangle = \int d^3r_1 \left[\sum_\alpha \left(\frac{Z_\alpha e^2 A^2(\vec{r}_1)}{4\pi\varepsilon_o |\vec{R}_\alpha - \vec{r}_i|}\right)\right] \quad (C3)$$

Similar expression is obtained for the external potential which gives:
$$\langle V_{ext}\rangle = \int d^3r_1..\int d^3r_N \left[\psi^*(\vec{r}_1,\vec{r}_2,...\vec{r}_N,t)\sum_i V_{ext}(\vec{r}_i)\psi(\vec{r}_1,\vec{r}_2,...\vec{r}_N,t)\right] \quad (C4)$$
which is reduced in the same way to:
$$\langle V_{ext}\rangle = \int d^3r_1 \rho(\vec{r}_1)V(\vec{r}_1) \quad (C5)$$
That again depends on the density (amplitude ) only.
$$\langle V_{ext}[A]\rangle = \int d^3r_1 A^2(\vec{r}_1)V(\vec{r}_1) \quad (C6)$$